\documentclass[a4paper,11pt]{article}
\usepackage{pos}
\usepackage{tikz}
\usepackage{xcolor}
\definecolor{plotred}{rgb}{0.8941176470588236, 0.10196078431372549, 0.10980392156862745}
\hypersetup{%
	pdftitle    = {Precision B*Bpi coupling from three-flavor lattice QCD},
	pdfauthor   = {A. Gérardin, J. Heitger, S. Kuberski, H. Simma, R. Sommer},
	pdfkeywords = {Lattice QCD, Heavy Quark Effective Theory, Bottom quarks, Heavy Meson chiral Effective Theory}
}%

\title{Precision $B^*B\pi$ coupling from three-flavor lattice QCD}

\renewcommand{\printHeadAuthors}{Simon Kuberski}
\manuallySeparateAuthors

\author[a]{Antoine G\'{e}rardin}
\author[b]{, Jochen Heitger}
\author*[c,d]{, Simon Kuberski}
\author[e]{, Hubert Simma}
\author[e, f]{\\ and Rainer Sommer}
\author{\\[.3cm]
	\begin{minipage}[b]{0.4\linewidth}
		\includegraphics[height=2.5\baselineskip]{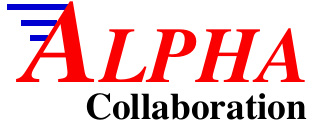}
	\end{minipage}
}
\affiliation[a]{Aix Marseille Univ, Universit\'{e} de Toulon, CNRS, CPT,\\ Marseille, France}

\affiliation[b]{Institut für Theoretische Physik, Westfälische Wilhelms-Universität Münster,\\	Wilhelm-Klemm-Straße 9, 48149 Münster, Germany}

\affiliation[c]{Helmholtz-Institut Mainz, Johannes Gutenberg-Universität Mainz,\\ Staudingerweg 18, 55128 Mainz, Germany}

\affiliation[d]{GSI Helmholtzzentrum für Schwerionenforschung,\\ Planckstraße 1, 64291 Darmstadt, Germany}

\affiliation[e]{John von Neumann Institute for Computing (NIC), DESY,\\
Platanenallee 6, 15738 Zeuthen, Germany}

\affiliation[f]{Institut für Physik, Humboldt-Universität zu Berlin,\\
Newtonstraße 15, 12489 Berlin, Germany}

\emailAdd{simon.kuberski@uni-mainz.de}

\abstract{We consider three-flavor QCD and perform a determination of the low-energy coupling $\hat{g}_\chi$ of SU(2) Heavy Meson Chiral Perturbation Theory. It is the $B^*B\pi$ coupling in the limit of static heavy and chiral light quarks and has not been determined with precision thus far. The calculation is performed on a large set of the $2+1$ flavor CLS ensembles with pion masses from 420\,MeV down to 130\,MeV.
This allows us to significantly reduce the systematic uncertainty from the chiral extrapolation compared to previous works. Only a weak dependence on the lattice spacing is visible in our results.\\[.5cm]

{
\hfill MS-TP-21-31\linebreak
\phantom{.}\hfill MITP-21-052\linebreak
\phantom{.}\hfill DESY-21-176\linebreak
}
}

\FullConference{%
 The 38th International Symposium on Lattice Field Theory, LATTICE2021
  26th-30th July, 2021
  Zoom/Gather@Massachusetts Institute of Technology
}

\begin{document}
\maketitle

\section{The \texorpdfstring{$B^*B\pi$}{B*Bpi} coupling in the static approximation}
The interactions of heavy-light mesons and soft pseudo-Goldstone bosons in Heavy Meson Chiral Perturbation Theory (HM$\chi$PT) at lowest order
are governed by a single low-energy constant $\hat{g}_\chi$ \cite{Casalbuoni:1996pg}. This constant is related to the would-be matrix element of the strong decay $B^* \rightarrow B\pi$ which is kinematically forbidden in nature and therefore inaccessible from experiment. In contrast, it is possible to compute the coupling $\hat{g}$ on the lattice \cite{deDivitiis:1998kj} from the matrix element of the light-light axial current,
\begin{align}
	\hat{g} = \frac{1}{2} \langle B^0(\textbf{0})| \hat{A}_k(0) |B^{*\dagger}_k(\textbf{0})\rangle\,,
\end{align}
in the limit of infinitely heavy b-quark mass. The calculation is performed using static heavy quarks and light quarks heavier than physical, followed by an extrapolation to vanishing light quark masses to obtain $\hat{g}_\chi$.

\begin{figure}
	\centering
	\includegraphics[width=.48\textwidth]{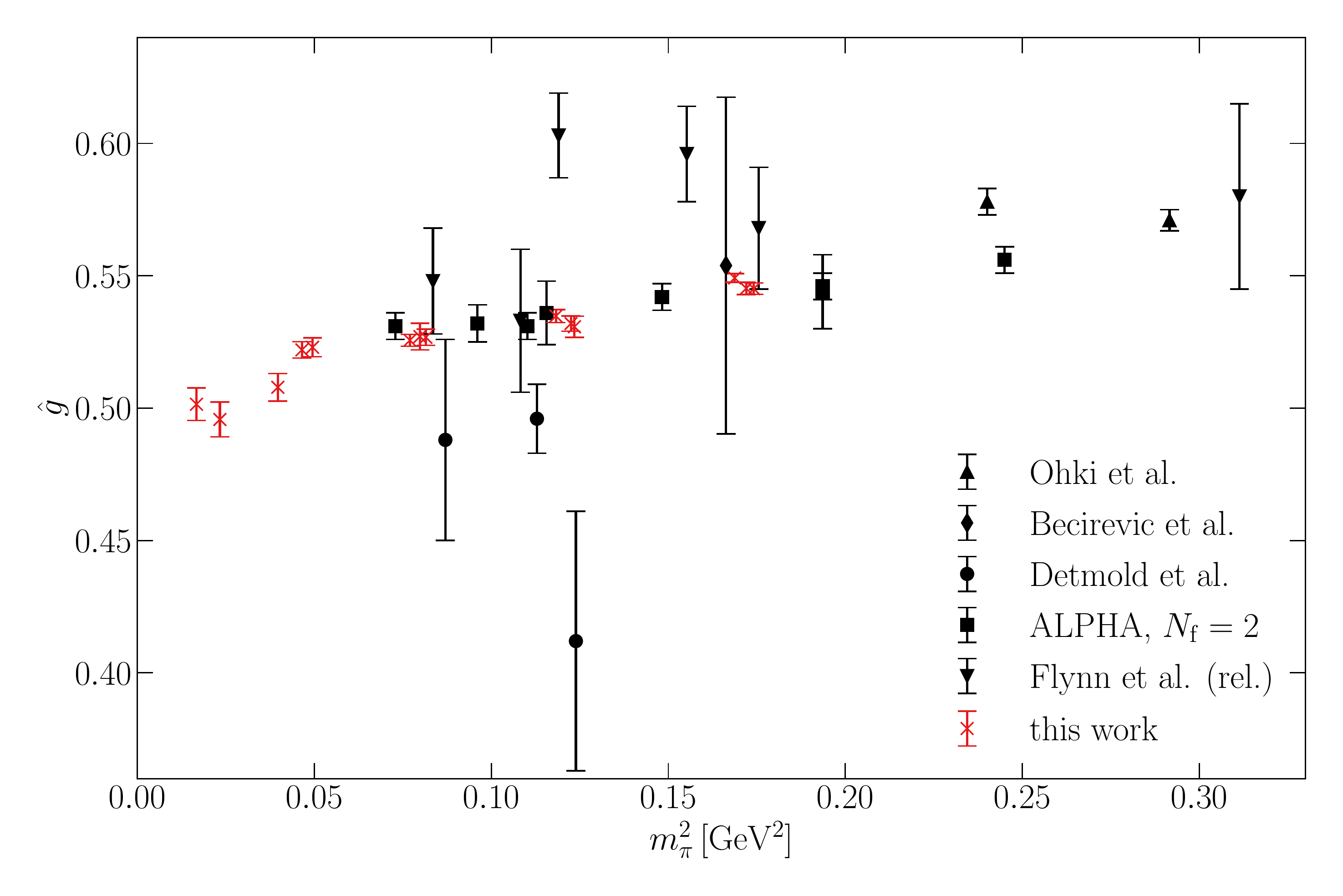}
	\includegraphics[width=.48\textwidth]{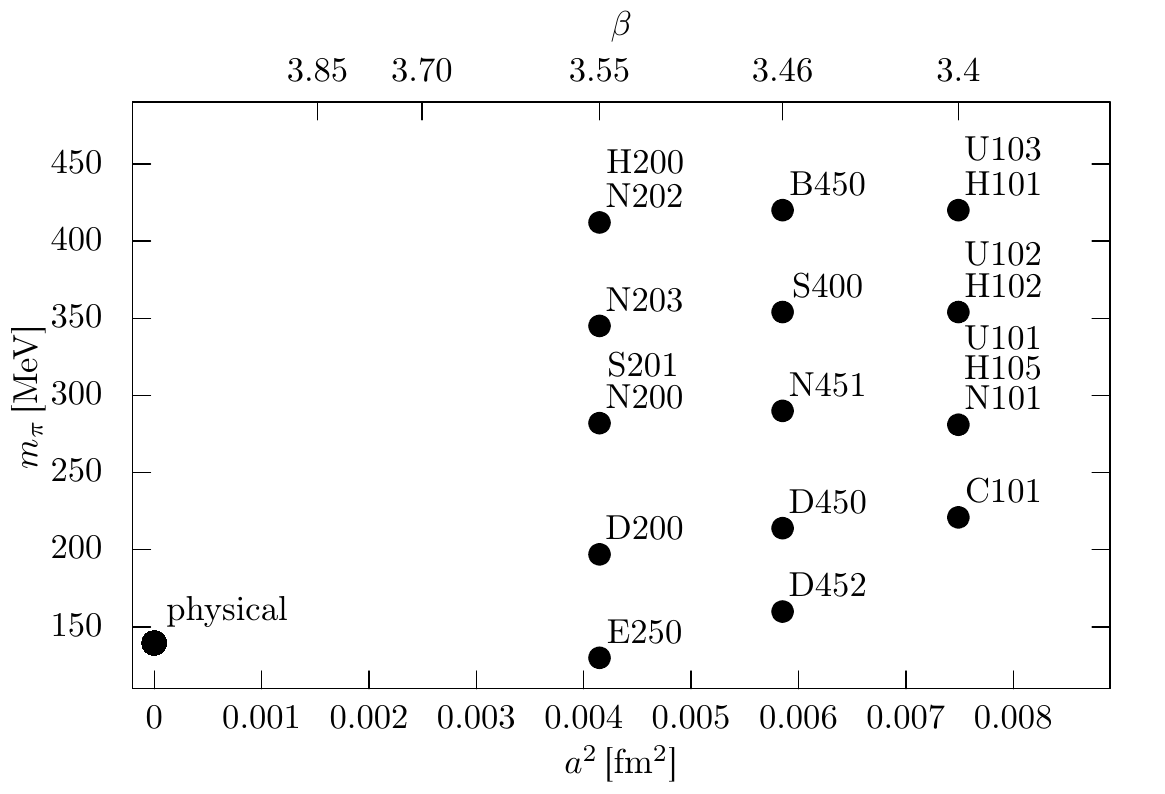}
	\caption{\textit{Left:} Dependence of $\hat{g}$ on the pion mass. Comparison of this work (in red) with existing work based on dynamical light quarks in Refs.~\cite{Ohki:2008py,Becirevic:2009yb,Detmold:2012ge,Bernardoni:2014kla,Flynn:2015xna}. The results have been obtained at finite lattice spacing. \textit{Right:} Overview of the ensembles used in this work in the landscape of squared lattice spacing and pion mass. The strange quark mass is implicitly fixed by the chiral trajectory. Multiple IDs at the same point indicate a variation of the spatial lattice size.\label{f:overview}}
\end{figure}

Amongst other applications, the precise knowledge of $\hat{g}_\chi$ is relevant to constrain chiral extrapolations of interesting $B$ physics observables that are computed on the lattice. In existing work using dynamical light quarks, a significant systematic uncertainty is present due to a long chiral extrapolation from pion masses above $270\,$MeV. This is illustrated on the left hand side of figure~\ref{f:overview} showing the results of Refs.~\cite{Ohki:2008py,Becirevic:2009yb,Detmold:2012ge,Bernardoni:2014kla,Flynn:2015xna} together with preliminary results of this work. In addition, systematic uncertainties due to excited-state contributions in the extraction of hadronic matrix elements require special attention in the analysis.

\section{Computing \texorpdfstring{$\hat{g}$}{g} on the lattice}
The extraction of matrix elements from lattice data is complicated by statistical errors that become significant for source-sink separations of $\mathrm{O}(1\,\mathrm{fm})$ and systematic errors due to the contamination by excited states at small distances. In order to obtain a reliable estimate for the matrix elements, both sources of uncertainty have to be reduced simultaneously as much as possible.
The suppression of excited-state contaminations in the extraction of hadron-to-hadron transition matrix elements has been discussed in Ref.~\cite{Bulava:2011yz}. There, a combination of the summation method \cite{Maiani:1987by} with the solution of a Generalized Eigenvalue Problem (GEVP) \cite{Michael:1985ne, Luscher:1990ck,Blossier:2009kd} has been introduced and shown to minimize systematic uncertainties, when compared to the summation method or the GEVP alone. 

The method is based on the computation of an $N\times N$ matrix of three-point functions 
\begin{align}
(D_{ij}^{3\mathrm{pt}})_k(t) &= \sum_{t_1=0}^{T-1} \langle \left(B^{*}_i\right)_k(t)\, (A_\mathrm{R})_k(t_1)\, B_j^\dagger(0) \rangle\,, \label{e:D3pt}
\end{align}
summed over the insertion time $t_1$. The correlation functions are constructed out of a set of $N$ interpolating operators, denoted by $i,j$ in eq.~(\ref{e:D3pt}). The index $k$ indicates the spatial polarization of the axial current. Together with the summed three-point function, the corresponding matrix of two-point functions 
\begin{align}
	C^\mathrm{2pt}_{ij}(t) = \langle B_i(t)B_j^\dagger(0) \rangle \label{e:C2pt}
\end{align}
is computed. The solution of a GEVP at time separations $t$ and $t_0$,
\begin{align}
C^\mathrm{2pt}(t)v_n(t, t_0) = \lambda_n(t, t_0) C^\mathrm{2pt}(t_0) v_n(t, t_0)\,,
\end{align}
gives access to the eigenvectors $v_n$ and eigenvalues $\lambda_n$, corresponding to the energies of the low-lying states \cite{Blossier:2009kd}. Following Ref.~\cite{Bulava:2011yz}, these can be employed to extract effective matrix elements via
\begin{align}
M^\mathrm{eff}_n(t, t_0) =
-\frac{1}{2} \partial_t \frac{\left(v_n(t, t_0), \left[\frac{D^\mathrm{3pt}(t) }{\lambda_n(t, t_0)} - D^\mathrm{3pt}(t_0)\right]v_n(t, t_0)\right)}{\left(v_n(t, t_0), C^\mathrm{2pt}(t_0) v_n(t, t_0)\right)}
= \hat{g}_{nn}+\mathrm{O}(e^{-(E_{N+1}-E_n){t}})\,. \label{e:Meff}
\end{align}
The effective matrix element $M_n^\mathrm{eff}$ matches $\hat{g}_{nn}$ up to corrections that vanish exponentially with the source-sink separation and the energy gap $\Delta_{N, n}\equiv(E_{N+1}-E_n)$ for a GEVP of size $N$, provided that $t_0\geq t/2$ is chosen. For the phenomenologically most relevant case of $n=1$, $\hat{g} \equiv g_{11}$, excited states are suppressed with $\Delta_{N, 1} > 1\,\mathrm{GeV}$ for $N>3$ in our calculation. The set of interpolating operators is constructed from smeared quark fields. We employ Gaussian smearing with APE-smeared gauge links \cite{Gusken:1989ad,Albanese:1987ds,Basak:2005gi}, as well as smearing via three-dimensional scalar and spinor auxiliary fields \cite{Papinutto:2018ajw}, and test different combinations of operators in our variational basis.
We note that it might be possible that the contamination by multi-hadron states, similar to those discussed in \cite{Bar:2018xyi}, is not entirely controlled by the used
GEVP. Worries are the dense spectrum of multi-hadron states in large volume and insufficient overlaps with those states when using just 
local interpolating fields.

To improve the signal-to-noise-ratio, we employ time-diluted stochastic sources \cite{Sommer:1994gg,Foley:2005ac} for the light quarks to profit from time-slice volume averaging. Three-point functions are obtained from sequential propagators. The use of HYP-smeared static quark actions \cite{DellaMorte:2003mn,DellaMorte:2005nwx} reduces statistical fluctuations by a factor that grows exponentially with the time separation. We employ the HYP1 and HYP2 actions of Ref.~\cite{DellaMorte:2005nwx} and obtain two sets of results that differ by $\mathrm{O}(a^2)$ effects.

\subsection{Computational Setup}
We perform our calculation at three resolutions on the $\mathrm{Tr}[M_\mathrm{q}]=\mathrm{const.}$ trajectory of the $N_\mathrm{f} = 2+1$ CLS ensembles \cite{Bruno:2014jqa}, generated with $\mathrm{O}(a)$ clover-improved Wilson quarks and tree-level improved Lüscher-Weisz gluons. An overview is given on the right hand side of figure \ref{f:overview}. Ensembles with open and periodic boundary conditions in time direction enter our analysis. We cover a range of light quark masses that spans from the $\mathrm{SU}(3)$-flavor-symmetric point, where $m_\pi =m_K \approx 420\,$MeV, down to slightly smaller than physical light quark masses. The strange quark mass is fixed implicitly by our chiral trajectory. At five points in the $(a^2,m_\pi)$ plane, we vary the spatial box sizes to explicitly test for finite-volume effects in our calculation.

Our quark action is $\mathrm{O}(a)$ improved using the determination of $c_\mathrm{SW}$ from Ref.~\cite{Bulava:2013cta}. The improvement and renormalization of the axial current that enters the effective matrix elements amounts to the renormalization pattern \cite{Bhattacharya:2005rb}
\begin{align}
(M_\mathrm{R}^\mathrm{eff})_{n} = Z_\mathrm{A} (1+b_\mathrm{A} am_\mathrm{q} + \bar{b}_\mathrm{A}a\mathrm{Tr}\left[M_\mathrm{q}\right])M_n^\mathrm{eff}\,.
\end{align}
We employ the renormalization constant $Z_\mathrm{A}$ that has been determined to high precision in \cite{DallaBrida:2018tpn} and the improvement coefficients $b_\mathrm{A}$ and  $\bar{b}_\mathrm{A}$ from Refs.~\cite{Bali:2021qem,Bali:toappear}. The critical hopping parameters that enter the computation of the bare subtracted quark masses $am_\mathrm{q}$ have been obtained from Refs.~\cite{Bali:2016umi,Gerardin:2018kpy}.

\begin{figure}
	\centering
	\includegraphics[width=.75\textwidth]{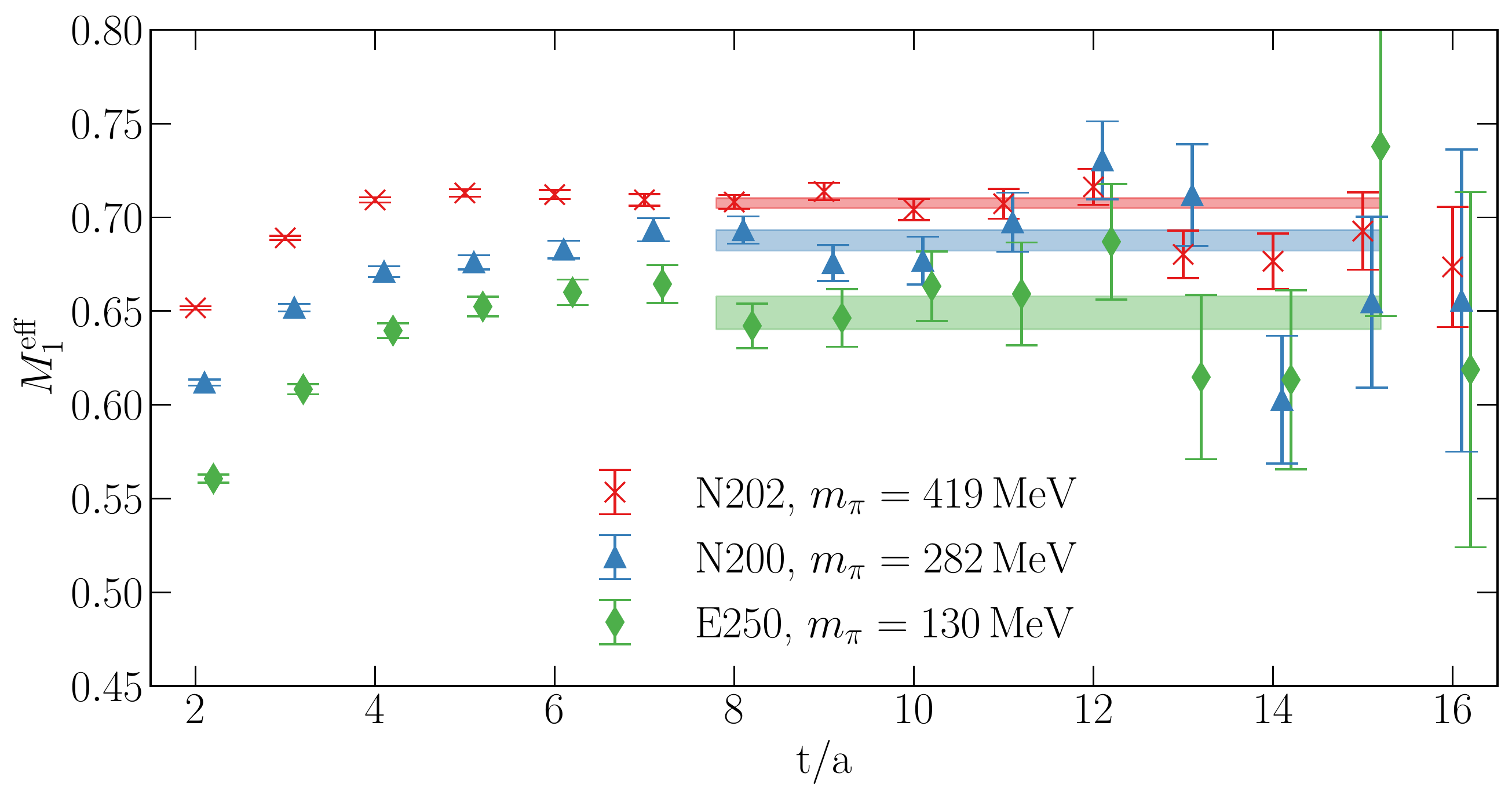}
	\caption{Representative extraction of the bare matrix elements $M^\mathrm{eff}_1$ at $a\approx 0.064\,\mathrm{fm}$ at three different pion masses, including the SU(3) flavor symmetric point and close to physical quark masses. \label{f:plateaus}}
\end{figure}

When solving the GEVP, we choose $t_0=t/2$ to suppress excited-state contamination, cf.~eq.~(\ref{e:Meff}). We determine the minimal time separation where excited state effects have sufficiently vanished by requiring 
\begin{align}
	\left|M^\mathrm{eff}_n(t)-M^\mathrm{eff}_n(t-\delta t)\right| < \sigma(t) \label{e:platcrit}
\end{align}
where $\delta t = \frac{1}{\Delta_{N, n}}$ is extracted from the GEVP via $\lambda_n(t)$ and $\sigma(t)$ are the statistical errors on $M^\mathrm{eff}_n(t)$. We find that the criterion of eq.~(\ref{e:platcrit}) is fulfilled for $t\geq 0.5\,$fm on all of our ensembles. To avoid a propagation of statistical fluctuations at single source-sink separations on some ensembles into the plateau averages, we fix $t_\mathrm{min}=0.52\,$fm for all ensembles. In a later stage of our analysis, we verify that a variation of $t_\mathrm{min}$ in a range $[0.3\,\mathrm{fm}, 0.9\,\mathrm{fm}]$ does not have a significant influence on the result in the chiral-continuum limit, albeit with large statistical uncertainties for $t_\mathrm{min}$ as large as $0.9\,$fm. We set the upper end of the plateau range to the time slice where the relative error on $M^\mathrm{eff}_n(t)$ exceeds $30\,\%$, which is the case at about $1\,$fm for the summed GEVP across all of our ensembles. In figure~\ref{f:plateaus}, we illustrate plateau fits at our finest lattice spacing at three different pion masses, including the physical one. It is visible that the signal deteriorates quickly. However, we are able to identify reasonable plateaus.

\section{Extrapolation to the chiral-continuum limit}
To arrive at $\hat{g}_\chi$, we have to extrapolate to the infinite volume, continuum and chiral limits. Our parametrization for the combined extrapolation is based on Ref.~\cite{Detmold:2011rb} and reads
\begin{align}
\hat{g}_{11}
\equiv \hat{g}_\chi\Bigg[&1
-(1+2\hat{g}_\chi^2)\, y \log y + c_1 y + c_2 y^2 + \dots \qquad & \text{\textit{chiral dependence}} \nonumber\\
&+ \hat{g}_\chi^2\, y \,F_0(m_\pi L) + y \, F_1(m_\pi L) + \dots & \text{\textit{finite-volume effects}}\\
&+ c_a a^2 + \dots \Bigg]  & \text{\textit{cutoff effects}} \nonumber\,,\\
\text{with}& \quad F_n(z)=
\mathrm{O}\left(\mathrm{e}^{-z} z^{-1/2-n}\right)\,,& \nonumber
\end{align}
where the dependence on the pion mass is parametrized via $y\equiv\frac{m_\pi^2}{8 \pi^2 f_\pi^2}$. A leading logarithmic dependence on $y$ is predicted from chiral perturbation theory and complicates the extrapolation from larger-than-physical pion masses to the chiral limit. The exact form of the leading finite-volume effects based on chiral perturbation theory has been derived in Ref.~\cite{Detmold:2011rb}. Since we work with $m_\pi L \gtrsim 4$, these effects are small: 
The leading $\chi$PT expression predicts a relative deviation of $0.5$\% at the symmetric point and sub per mill effects at physical pion mass for $m_\pi L = 4$. Due to the non-perturbative $\mathrm{O}(a)$ improvement of action and currents, the leading cutoff effects are of $\mathrm{O}(a^2)$.

\begin{figure}
	\centering
	\begin{tikzpicture}
	\draw (0, 0) node[inner sep=0] {\includegraphics[width=.95\textwidth]{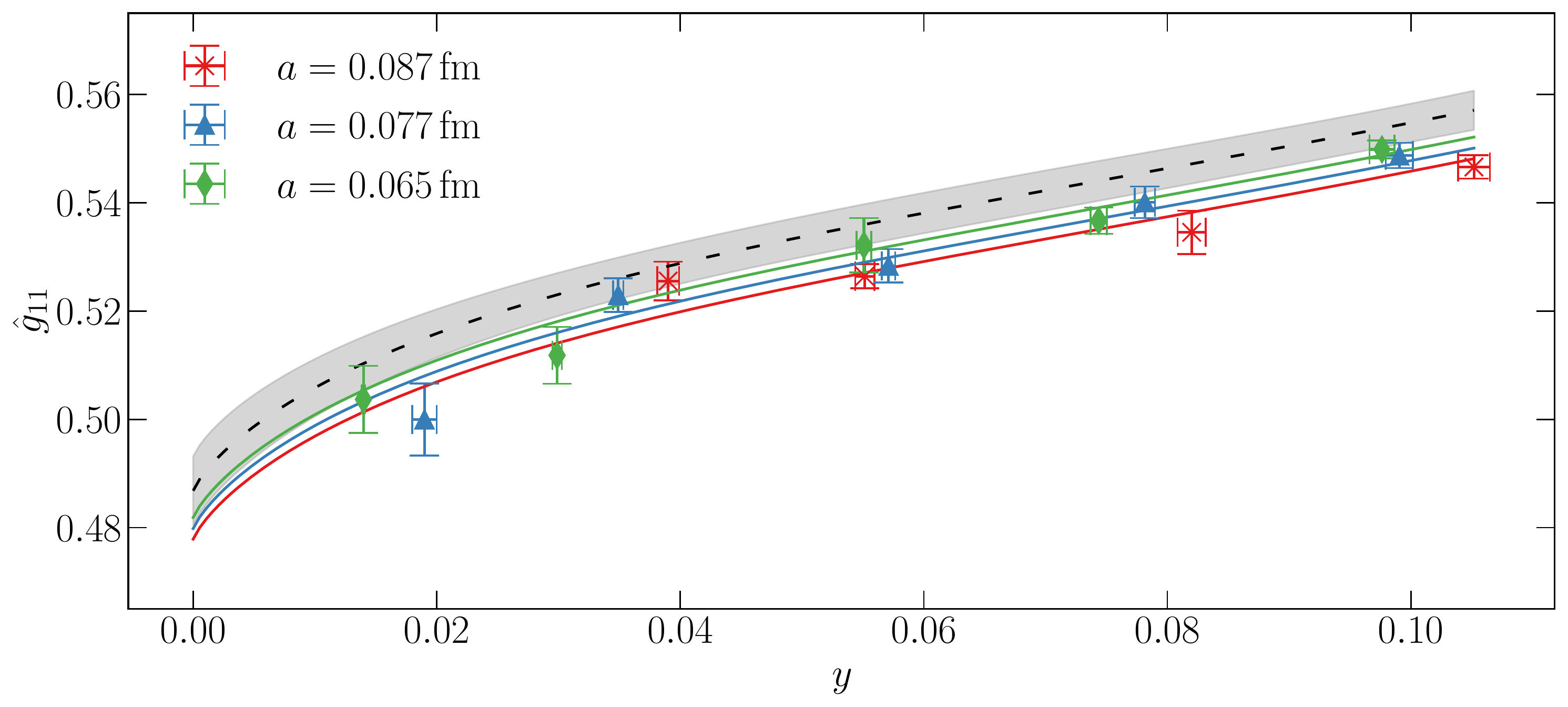}};
	\draw (4.5,-1.5) node {\textcolor{plotred}{Preliminary}};
	\end{tikzpicture}
	\caption{Illustration of a preliminary combined chiral and continuum extrapolation. Results at three lattice spacings enter the fit, as indicated by different colors. The corresponding lines show the quark mass dependence at those lattice spacings, as determined by the fit. The gray band gives the uncertainty on the quark mass dependence in the continuum limit. We show results using the HYP2 action for the static quarks. \label{f:cc}}
\end{figure}

We carry out a combined fit to extract $\hat{g}_\chi$ as our estimate for the (infinite-volume) LEC of HM$\chi$PT. Figure \ref{f:cc} illustrates an exemplary fit to the renormalized effective matrix elements. We show the results at finite lattice spacing together with the fit function, evaluated at all three resolutions (colored lines), as well as the functional form in the continuum limit. Here, the statistical uncertainty is depicted by the gray band. The inclusion of a term proportional to $y^2$ allows us to describe the data in the complete range of pion masses. Despite the inclusion of results close to physical pion masses, it is hard to constrain the chiral limit due to the presence of the logarithmic term in the extrapolation. However, for the first time, a deviation from a linear behavior in $y$ towards the chiral limit is visible in our data. The systematic uncertainty due to the chiral extrapolation, which has to be included in our final results, will be estimated based on a variation of the functional form and cuts on the maximal pion mass.

Cutoff effects appear to be small for both static quark discretizations. Nevertheless, we are able to resolve a dependence on $a^2$ for both actions. Our continuum extrapolations of results based on HYP1 and HYP2 actions coincide with each other and with a combined extrapolation of the two sets. The data is compatible with cutoff effects proportional to $a^2$ without a mass-dependence.

\begin{figure}
	\centering
	\includegraphics[width=.48\textwidth]{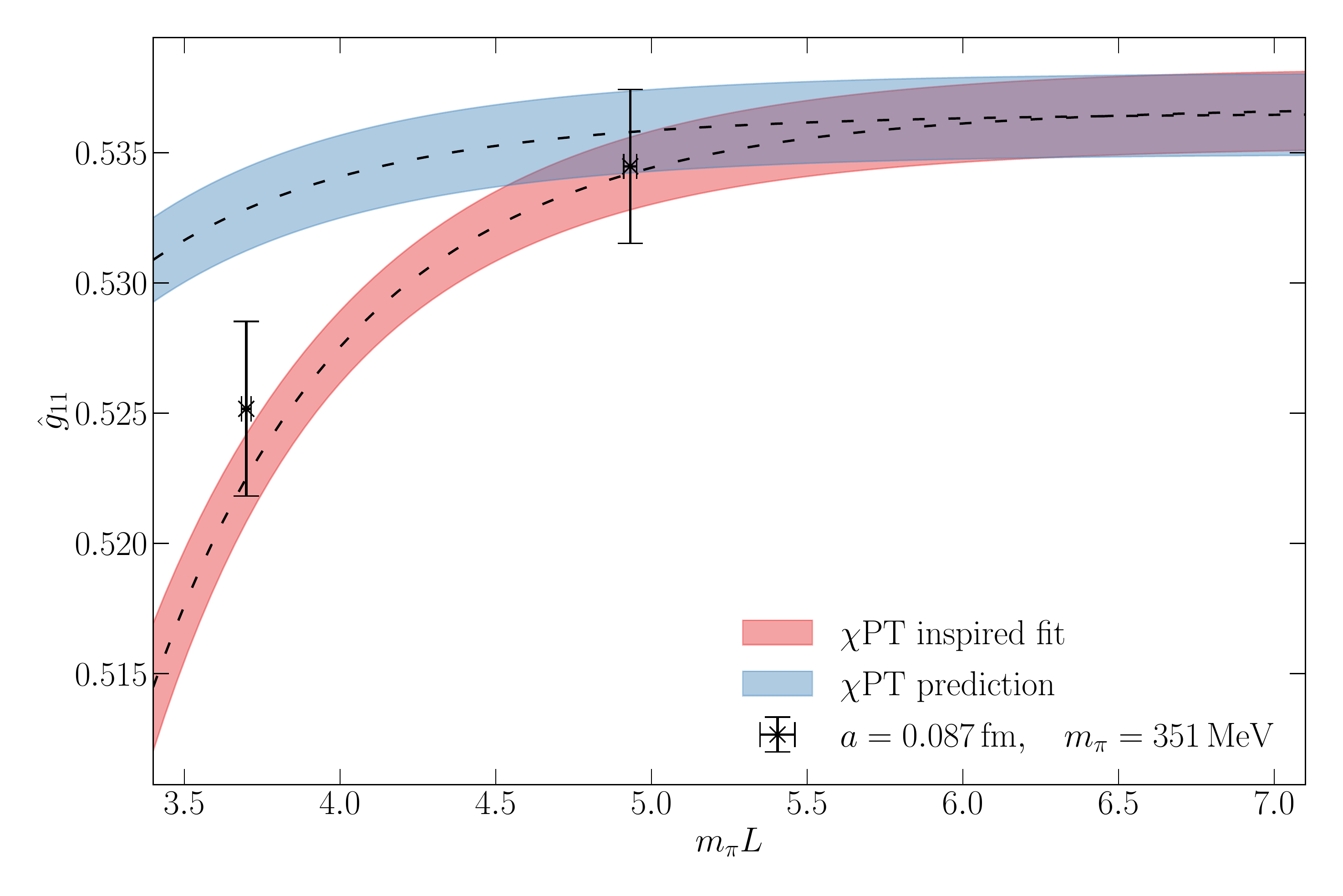}
	\includegraphics[width=.48\textwidth]{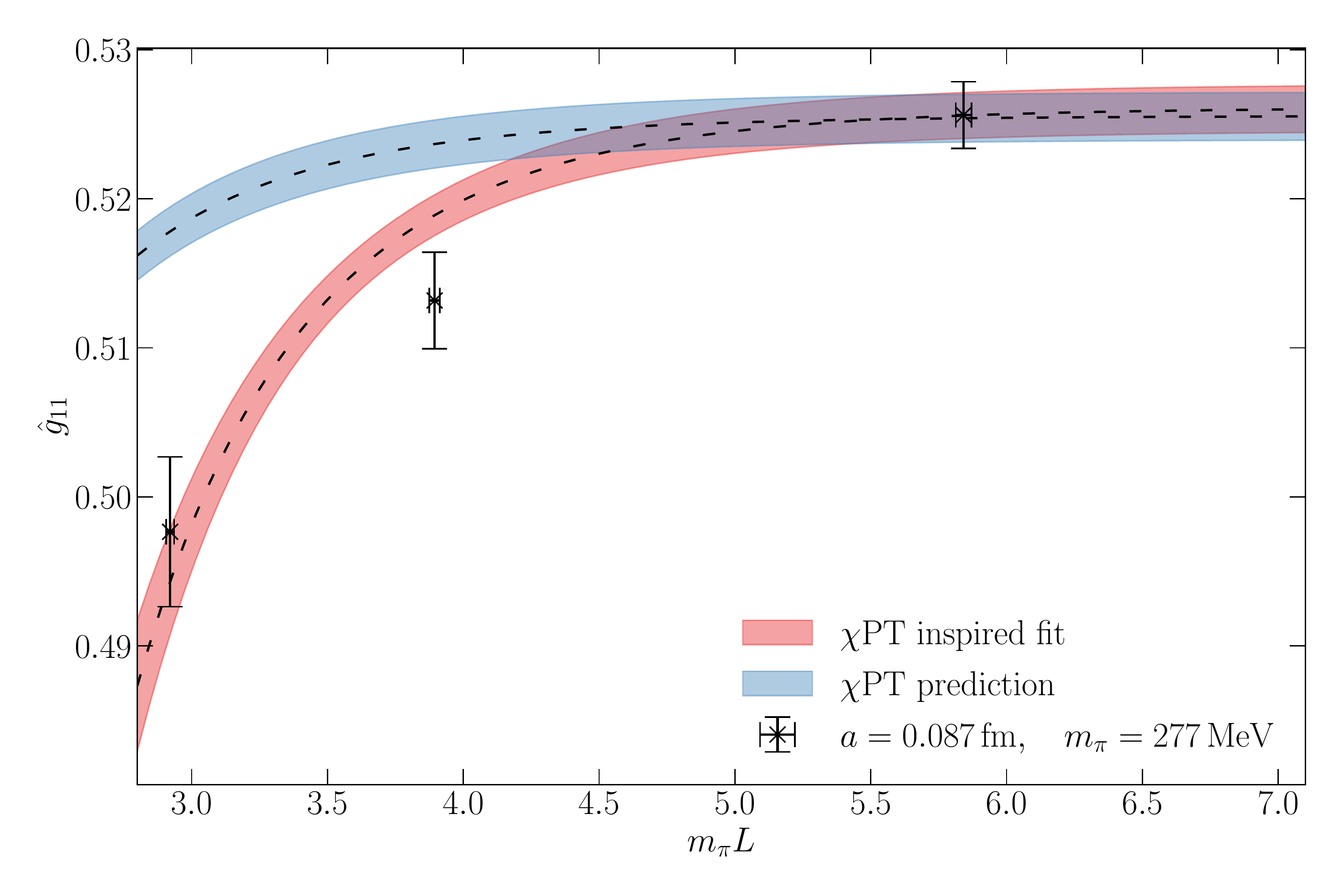}
	\caption{Finite-volume effects for two different pion masses at the coarsest lattice spacing in our analysis. We show a comparison of fit results using a free fit parameter ${c_\mathrm{FV}}$ and using ${c_\mathrm{FV}}=1$ in the combined chiral, continuum and finite-volume extrapolation. \label{f:fv}}
\end{figure}

We adopt and compare different strategies for the description of finite-volume effects. Based on the ensembles with reduced box sizes, which are shown in figure~\ref{f:overview}, we are able to compare the expectation from chiral perturbation theory with our data and to include the observed deviations in our fits. In figure~\ref{f:fv} we illustrate the renormalized effective matrix elements as determined from plateau fits at two different pion masses and $a\approx0.087\,$fm. Together with the data, we show the NLO prediction based on chiral perturbation theory. It is apparent that this prediction is not able to describe the deviations from the infinite-volume limit at small values of $m_\pi L$. However, these effects appear to be small for the larger boxes. 
For comparison, we also show the result of a fit to the finite-volume effects, which is obtained by including the small volumes in a combined chiral-continuum-volume extrapolation. For this fit, we add a fit parameter ${c_\mathrm{FV}}$ to the analytical form, 
\begin{align}
\hat{g}_{11}^\mathrm{FV} = \hat{g}_\chi \left[1 + \hat{g}_\chi^2\, y \,F_0(m_\pi L) + {c_\mathrm{FV}}y \, F_1(m_\pi L)\right]\,,
\end{align}
where ${c_\mathrm{FV}}=1$ in $\chi$PT. We are able to describe the data at small $m_\pi L$ with the fit result. At large $m_\pi L$, the deviation from the $\chi$PT prediction is small. Consequently, the extrapolated result does not change significantly, when we switch between the two descriptions, or even neglect finite-volume effects at all.

\section{Conclusions}
We will be able to determine $\hat{g}_\chi$ with much improved precision and better controlled systematics compared to existing work. We employ a number of well-established techniques, such as the use of random sources and the summed GEVP to extend the window where an extraction of matrix elements is possible without the introduction of large statistical and systematic uncertainties. The main improvement of our calculation with respect to existing work is the inclusion of a number of ensembles at pion masses below $270\,$MeV down to $130\,$MeV. This allows us to constrain the chiral extrapolation more tightly and to reduce the leading systematic uncertainty that was present in the so far most precise lattice calculation of Ref.~\cite{Bernardoni:2014kla}. The final assessment of the remaining systematic uncertainty is not yet finished and will be addressed in the forthcoming publication \cite{Gerardin:toappear}.
The knowledge of $\hat{g}_\chi$ will help us to perform chiral extrapolations of phenomenologically interesting quantities in the ALPHA program for $B$ physics with $2+1$ flavors of light quarks \cite{Heitger:2003nj,DellaMorte:2013ega,Fritzsch:2018yag}. 

\subsection*{Acknowledgements}
The work of JH and SK has been supported by the Deutsche Forschungsgemeinschaft (DFG) through the Research Training Group {\it GRK 2149: Strong and Weak Interactions – from Hadrons to Dark Matter}. This publication received funding from the Excellence Initiative of Aix-Marseille University -- A*MIDEX, a French “Investissements d’Avenir” programme, AMX-18-ACE-005. The authors gratefully acknowledge the Gauss Centre for Supercomputing e.V. (\url{www.gauss-centre.eu}) for funding this project by providing computing time on the GCS Supercomputer SuperMUC-NG at Leibniz Supercomputing Centre (\url{www.lrz.de}). We furthermore acknowledge the computer resources provided by the WWU IT of the University of Münster (PALMA-II HPC cluster) and by DESY Zeuthen (PAX cluster) and thank the staff of the computing centers for their support. We are grateful to our colleagues in the CLS initiative for producing the gauge field configuration ensembles used in this study.

\bibliographystyle{JHEP}
\bibliography{bibliography}

\providecommand{\href}[2]{#2}\begingroup\raggedright\begin{thebibliography}{10}

\bibitem{Casalbuoni:1996pg}
R.~Casalbuoni, A.~Deandrea, N.~Di~Bartolomeo, R.~Gatto, F.~Feruglio and
  G.~Nardulli, \emph{{Phenomenology of heavy meson chiral Lagrangians}},
  \href{https://doi.org/10.1016/S0370-1573(96)00027-0}{\emph{Phys. Rept.}
  {\bfseries 281} (1997) 145}
  [\href{https://arxiv.org/abs/hep-ph/9605342}{{\ttfamily hep-ph/9605342}}].

\bibitem{deDivitiis:1998kj}
G.M.~de~Divitiis, L.~Del~Debbio, M.~Di~Pierro, J.M.~Flynn, C.~Michael and
  J.~Peisa, \emph{{Towards a lattice determination of the $B^*B\pi$ coupling}},
  {\emph{JHEP} {\bfseries 10} (1998) 010}
  [\href{https://arxiv.org/abs/hep-lat/9807032}{{\ttfamily hep-lat/9807032}}].

\bibitem{Ohki:2008py}
H.~Ohki, H.~Matsufuru and T.~Onogi, \emph{{Determination of $B^*B\pi$ coupling
  in unquenched QCD}},
  \href{https://doi.org/10.1103/PhysRevD.77.094509}{\emph{Phys. Rev. D}
  {\bfseries 77} (2008) 094509}
  [\href{https://arxiv.org/abs/0802.1563}{{\ttfamily 0802.1563}}].

\bibitem{Becirevic:2009yb}
D.~Becirevic, B.~Blossier, E.~Chang and B.~Haas, \emph{{$g_{B^*B\pi}$-coupling
  in the static heavy quark limit}},
  \href{https://doi.org/10.1016/j.physletb.2009.07.031}{\emph{Phys. Lett. B}
  {\bfseries 679} (2009) 231}
  [\href{https://arxiv.org/abs/0905.3355}{{\ttfamily 0905.3355}}].

\bibitem{Detmold:2012ge}
W.~Detmold, C.J.D.~Lin and S.~Meinel, \emph{{Calculation of the heavy-hadron
  axial couplings $g_1$, $g_2$ and $g_3$ using lattice QCD}},
  \href{https://doi.org/10.1103/PhysRevD.85.114508}{\emph{Phys. Rev. D}
  {\bfseries 85} (2012) 114508}
  [\href{https://arxiv.org/abs/1203.3378}{{\ttfamily 1203.3378}}].

\bibitem{Bernardoni:2014kla}
F.~Bernardoni, J.~Bulava, M.~Donnellan and R.~Sommer, \emph{{Precision lattice
  QCD computation of the $B^*B\pi$ coupling}},
  \href{https://doi.org/10.1016/j.physletb.2014.11.051}{\emph{Phys. Lett. B}
  {\bfseries 740} (2015) 278}
  [\href{https://arxiv.org/abs/1404.6951}{{\ttfamily 1404.6951}}].

\bibitem{Flynn:2015xna}
J.M.~Flynn, P.~Fritzsch, T.~Kawanai, C.~Lehner, B.~Samways, C.T.~Sachrajda
  et~al., \emph{{The $B^*B\pi$ Coupling Using Relativistic Heavy Quarks}},
  \href{https://doi.org/10.1103/PhysRevD.93.014510}{\emph{Phys. Rev. D}
  {\bfseries 93} (2016) 014510}
  [\href{https://arxiv.org/abs/1506.06413}{{\ttfamily 1506.06413}}].

\bibitem{Bulava:2011yz}
J.~Bulava, M.~Donnellan and R.~Sommer, \emph{{On the computation of
  hadron-to-hadron transition matrix elements in lattice QCD}},
  \href{https://doi.org/10.1007/JHEP01(2012)140}{\emph{JHEP} {\bfseries 01}
  (2012) 140} [\href{https://arxiv.org/abs/1108.3774}{{\ttfamily 1108.3774}}].

\bibitem{Maiani:1987by}
L.~Maiani, G.~Martinelli, M.L.~Paciello and B.~Taglienti, \emph{{Scalar
  Densities and Baryon Mass Differences in Lattice {QCD} with Wilson
  Fermions}}, \href{https://doi.org/10.1016/0550-3213(87)90078-2}{\emph{Nucl.
  Phys. B} {\bfseries 293} (1987) 420}.

\bibitem{Michael:1985ne}
C.~Michael, \emph{{Adjoint Sources in Lattice Gauge Theory}},
  \href{https://doi.org/10.1016/0550-3213(85)90297-4}{\emph{Nucl. Phys.}
  {\bfseries B259} (1985) 58}.

\bibitem{Luscher:1990ck}
M.~Lüscher and U.~Wolff, \emph{{How to Calculate the Elastic Scattering Matrix
  in Two-dimensional Quantum Field Theories by Numerical Simulation}},
  \href{https://doi.org/10.1016/0550-3213(90)90540-T}{\emph{Nucl. Phys.}
  {\bfseries B339} (1990) 222}.

\bibitem{Blossier:2009kd}
B.~Blossier, M.~Della~Morte, G.~von Hippel, T.~Mendes and R.~Sommer, \emph{{On
  the generalized eigenvalue method for energies and matrix elements in lattice
  field theory}},
  \href{https://doi.org/10.1088/1126-6708/2009/04/094}{\emph{JHEP} {\bfseries
  04} (2009) 094} [\href{https://arxiv.org/abs/0902.1265}{{\ttfamily
  0902.1265}}].

\bibitem{Gusken:1989ad}
S.~Güsken, U.~Löw, K.H.~Mütter, R.~Sommer, A.~Patel and K.~Schilling,
  \emph{{Nonsinglet Axial Vector Couplings of the Baryon Octet in Lattice
  {QCD}}}, \href{https://doi.org/10.1016/S0370-2693(89)80034-6}{\emph{Phys.
  Lett.} {\bfseries B227} (1989) 266}.

\bibitem{Albanese:1987ds}
M.~Albanese et~al., \emph{{Glueball Masses and String Tension in Lattice QCD}},
  \href{https://doi.org/10.1016/0370-2693(87)91160-9}{\emph{Phys.\ Lett.\ B}
  {\bfseries 192} (1987) 163}.

\bibitem{Basak:2005gi}
S.~Basak, I.~Sato, S.~Wallace, R.~Edwards, D.~Richards, G.T.~Fleming et~al.,
  \emph{{Combining quark and link smearing to improve extended baryon
  operators}}, \href{https://doi.org/10.22323/1.020.0076}{\emph{PoS} {\bfseries
  LAT2005} (2006) 076} [\href{https://arxiv.org/abs/hep-lat/0509179}{{\ttfamily
  hep-lat/0509179}}].

\bibitem{Papinutto:2018ajw}
M.~Papinutto, F.~Scardino and S.~Schaefer, \emph{{New extended interpolating
  fields built from three-dimensional fermions}},
  \href{https://doi.org/10.1103/PhysRevD.98.094506}{\emph{Phys. Rev. D}
  {\bfseries 98} (2018) 094506}
  [\href{https://arxiv.org/abs/1807.08714}{{\ttfamily 1807.08714}}].

\bibitem{Bar:2018xyi}
O.~Bär, \emph{{$N\pi$-state contamination in lattice calculations of the
  nucleon axial form factors}},
  \href{https://doi.org/10.1103/PhysRevD.99.054506}{\emph{Phys. Rev. D}
  {\bfseries 99} (2019) 054506}
  [\href{https://arxiv.org/abs/1812.09191}{{\ttfamily 1812.09191}}].

\bibitem{Sommer:1994gg}
R.~Sommer, \emph{{Leptonic decays of B and D mesons}},
  \href{https://doi.org/10.1016/0920-5632(95)00201-J}{\emph{Nucl. Phys. B Proc.
  Suppl.} {\bfseries 42} (1995) 186}
  [\href{https://arxiv.org/abs/hep-lat/9411024}{{\ttfamily hep-lat/9411024}}].

\bibitem{Foley:2005ac}
J.~Foley, K.~Jimmy~Juge, A.~\'{O}~Cais, M.~Peardon, S.M.~Ryan and
  J.-I.~Skullerud, \emph{{Practical all-to-all propagators for lattice QCD}},
  \href{https://doi.org/10.1016/j.cpc.2005.06.008}{\emph{Comput. Phys. Commun.}
  {\bfseries 172} (2005) 145}
  [\href{https://arxiv.org/abs/hep-lat/0505023}{{\ttfamily hep-lat/0505023}}].

\bibitem{DellaMorte:2003mn}
M.~Della~Morte, S.~Dürr, J.~Heitger, H.~Molke, J.~Rolf, A.~Shindler et~al.,
  \emph{{Lattice HQET with exponentially improved statistical precision}},
  \href{https://doi.org/10.1016/j.physletb.2005.03.017}{\emph{Phys. Lett. B}
  {\bfseries 581} (2004) 93}
  [\href{https://arxiv.org/abs/hep-lat/0307021}{{\ttfamily hep-lat/0307021}}].

\bibitem{DellaMorte:2005nwx}
M.~Della~Morte, A.~Shindler and R.~Sommer, \emph{{On lattice actions for static
  quarks}}, \href{https://doi.org/10.1088/1126-6708/2005/08/051}{\emph{JHEP}
  {\bfseries 08} (2005) 051}
  [\href{https://arxiv.org/abs/hep-lat/0506008}{{\ttfamily hep-lat/0506008}}].

\bibitem{Bruno:2014jqa}
M.~Bruno et~al., \emph{{Simulation of QCD with N$_{f} =$ 2 $+$ 1 flavors of
  non-perturbatively improved Wilson fermions}},
  \href{https://doi.org/10.1007/JHEP02(2015)043}{\emph{JHEP} {\bfseries 02}
  (2015) 043} [\href{https://arxiv.org/abs/1411.3982}{{\ttfamily 1411.3982}}].

\bibitem{Bulava:2013cta}
J.~Bulava and S.~Schaefer, \emph{{Improvement of $N_f$ = 3 lattice QCD with
  Wilson fermions and tree-level improved gauge action}},
  \href{https://doi.org/10.1016/j.nuclphysb.2013.05.019}{\emph{Nucl. Phys. B}
  {\bfseries 874} (2013) 188}
  [\href{https://arxiv.org/abs/1304.7093}{{\ttfamily 1304.7093}}].

\bibitem{Bhattacharya:2005rb}
T.~Bhattacharya, R.~Gupta, W.~Lee, S.R.~Sharpe and J.M.S.~Wu, \emph{{Improved
  bilinears in lattice QCD with non-degenerate quarks}},
  \href{https://doi.org/10.1103/PhysRevD.73.034504}{\emph{Phys. Rev.}
  {\bfseries D73} (2006) 034504}
  [\href{https://arxiv.org/abs/hep-lat/0511014}{{\ttfamily hep-lat/0511014}}].

\bibitem{DallaBrida:2018tpn}
M.~Dalla~Brida, T.~Korzec, S.~Sint and P.~Vilaseca, \emph{{High precision
  renormalization of the flavour non-singlet Noether currents in lattice QCD
  with Wilson quarks}},
  \href{https://doi.org/10.1140/epjc/s10052-018-6514-5}{\emph{Eur. Phys. J. C}
  {\bfseries 79} (2019) 23} [\href{https://arxiv.org/abs/1808.09236}{{\ttfamily
  1808.09236}}].

\bibitem{Bali:2021qem}
G.S.~Bali, V.~Braun, S.~Collins, A.~Sch\"afer and J.~Simeth, \emph{{Masses and
  decay constants of the \ensuremath{\eta} and \ensuremath{\eta}' mesons from
  lattice QCD}}, \href{https://doi.org/10.1007/JHEP08(2021)137}{\emph{JHEP}
  {\bfseries 08} (2021) 137}
  [\href{https://arxiv.org/abs/2106.05398}{{\ttfamily 2106.05398}}].

\bibitem{Bali:toappear}
G.S.~Bali et~al., to appear.

\bibitem{Bali:2016umi}
G.S.~Bali, E.E.~Scholz, J.~Simeth and W.~S\"oldner, \emph{{Lattice simulations
  with $N_f=2+1$ improved Wilson fermions at a fixed strange quark mass}},
  \href{https://doi.org/10.1103/PhysRevD.94.074501}{\emph{Phys. Rev. D}
  {\bfseries 94} (2016) 074501}
  [\href{https://arxiv.org/abs/1606.09039}{{\ttfamily 1606.09039}}].

\bibitem{Gerardin:2018kpy}
A.~G\'erardin, T.~Harris and H.B.~Meyer, \emph{{Nonperturbative renormalization
  and $O(a)$-improvement of the nonsinglet vector current with $N_f=2+1$ Wilson
  fermions and tree-level Symanzik improved gauge action}},
  \href{https://doi.org/10.1103/PhysRevD.99.014519}{\emph{Phys. Rev. D}
  {\bfseries 99} (2019) 014519}
  [\href{https://arxiv.org/abs/1811.08209}{{\ttfamily 1811.08209}}].

\bibitem{Detmold:2011rb}
W.~Detmold, C.J.D.~Lin and S.~Meinel, \emph{{Axial couplings in heavy hadron
  chiral perturbation theory at the next-to-leading order}},
  \href{https://doi.org/10.1103/PhysRevD.84.094502}{\emph{Phys. Rev. D}
  {\bfseries 84} (2011) 094502}
  [\href{https://arxiv.org/abs/1108.5594}{{\ttfamily 1108.5594}}].

\bibitem{Gerardin:toappear}
A.~G\'erardin, J.~Heitger and S.~Kuberski, \emph{{The static $B^*B\pi$ coupling
  of three flavor QCD}},  in preparation.

\bibitem{Heitger:2003nj}
J.~Heitger and R.~Sommer, \emph{{Nonperturbative heavy quark effective
  theory}}, \href{https://doi.org/10.1088/1126-6708/2004/02/022}{\emph{JHEP}
  {\bfseries 02} (2004) 022}
  [\href{https://arxiv.org/abs/hep-lat/0310035}{{\ttfamily hep-lat/0310035}}].

\bibitem{DellaMorte:2013ega}
M.~Della~Morte, S.~Dooling, J.~Heitger, D.~Hesse and H.~Simma, \emph{{Matching
  of heavy-light flavour currents between HQET at order 1/$m$ and QCD: I.
  Strategy and tree-level study}},
  \href{https://doi.org/10.1007/JHEP05(2014)060}{\emph{JHEP} {\bfseries 05}
  (2014) 060} [\href{https://arxiv.org/abs/1312.1566}{{\ttfamily 1312.1566}}].

\bibitem{Fritzsch:2018yag}
P.~Fritzsch, J.~Heitger and S.~Kuberski, \emph{{$\mathcal{O}(a)$ improved quark
  mass renormalization for a non-perturbative matching of HQET to three-flavor
  QCD}}, \href{https://doi.org/10.22323/1.334.0218}{\emph{PoS} {\bfseries
  LATTICE2018} (2018) 218} [\href{https://arxiv.org/abs/1811.02591}{{\ttfamily
  1811.02591}}].

\end{thebibliography}\endgroup

\end{document}